\documentclass[aps,prb,twocolumn,bibliography,superscriptaddress]{revtex4-1}
\usepackage[utf8]{inputenc}
\usepackage{amsmath}
\usepackage{physics}
\usepackage{mathtools}
\usepackage{amsfonts}
\usepackage{amssymb}
\usepackage{graphicx}
\usepackage{subfigure}
\usepackage{textcomp}
\usepackage{color}
\usepackage[dvipsnames]{xcolor}

\begin{document}

\title{Transport properties of a non-Hermitian Weyl semimetal}

\author{Soumi Dey}
\email{soumidey@iisc.ac.in}
\affiliation{Solid State and Structural Chemistry Unit, Indian Institute of Science, Bangalore 560012, India}
\author{Ayan Banerjee}
\email{ayanbanerjee@iisc.ac.in}
\affiliation{Solid State and Structural Chemistry Unit, Indian Institute of Science, Bangalore 560012, India}
\author{Debashree Chowdhury}
\email{debashreephys@gmail.com}
\affiliation{Centre for Nanotechnology, Indian Institute of Technology Roorkee, Roorkee, Uttarakhand-247667}
\author{Awadhesh Narayan}
\email{awadhesh@iisc.ac.in}
\affiliation{Solid State and Structural Chemistry Unit, Indian Institute of Science, Bangalore 560012, India}

\date{\today}

\begin{abstract}
In recent years, non-Hermitian (NH) topological semimetals have garnered significant attention due to their unconventional properties. In this work, we explore the transport properties of a three-dimensional dissipative Weyl semi-metal formed as a result of the stacking of two-dimensional Chern insulators. We find that unlike Hermitian systems where the Hall conductance is quantized, in presence of non-Hermiticity, the quantized Hall conductance starts to deviate from its usual nature. We show that the non-quantized nature of the Hall conductance in such NH topological systems is intimately connected to the presence of exceptional points (EPs). We find that in the case of open boundary conditions, the transition from a topologically trivial regime to a non-trivial topological regime takes place at a different value of the momentum than that of the periodic boundary spectra. This discrepancy is solved by considering the non-Bloch case and the generalized Brillouin zone (GBZ). Finally, we present the Hall conductance evaluated over the GBZ and connect it to the separation between the Weyl nodes, within the non-Bloch theory.
\end{abstract}

\maketitle
 

\section{Introduction}
Quantum states with topological protection is a topic that has been at the forefront of research for some time now. In this arena of research, the classification of the topological phases arises due to different symmetries preserved or broken in the system~\cite{hasan2010colloquium}. The branch of topology started entering into condensed matter physics soon after the discovery of the quantum Hall effect~\cite{von1986quantized}. Subsequently, new materials with unique topological properties were discovered and coined as topological insulators (TIs)~\cite{kane2005quantum,kane2005z,moore2010birth,wen1995topological,roy2009z,hasan2010colloquium}, where the spin-orbit interaction plays a crucial role. In TIs, the existence of unusual edge properties is protected by time-reversal (TR) symmetry. On the other hand, in some cases, we encounter topologically protected edge states for TR symmetry broken two-dimensional (2D) systems, which are termed as Chern insulating phases~\cite{haldane1988model}.  Furthermore, topologically nontrivial phases also arise in gap-less materials. Similar to graphene in 2D~\cite{neto2009electronic}, Weyl semimetals (WSMs)~\cite{yang2011quantum,wan2011topological,lu2013weyl,xu2015discovery,lv2015observation,yang2015weyl,soluyanov2015type,xiao2015synthetic,lu2016symmetry,chen2016photonic,lin2016photonic,xiao2016hyperbolic,lu2015experimental,noh2017experimental} show a gap-less band structure in 3D. In WSMs the conduction and valence bands touch each other at some special points, i.e., the Weyl nodes, which appear in pairs with quantized Berry charge. The surface states of these systems are in the form of an open-ended arc, coined as the Fermi arc. These Weyl nodes are protected from several kinds of disorder, apart from those with broken discrete translation symmetry or broken charge conservation symmetry~\cite{yang2011quantum,wan2011topological,lu2013weyl,xu2015discovery,lv2015observation,yang2015weyl,soluyanov2015type,xiao2015synthetic,lu2016symmetry,chen2016photonic,lin2016photonic,xiao2016hyperbolic,lu2015experimental,noh2017experimental}. Remarkably, the fact that the WSM phases can be achieved by stacking multiple layers of Chern insulators has been discussed in literature~\cite{burkov2011weyl,yang2011quantum,shapourian2016phase}.

\begin{figure*}
    \centering
    \includegraphics[width=0.8\linewidth]{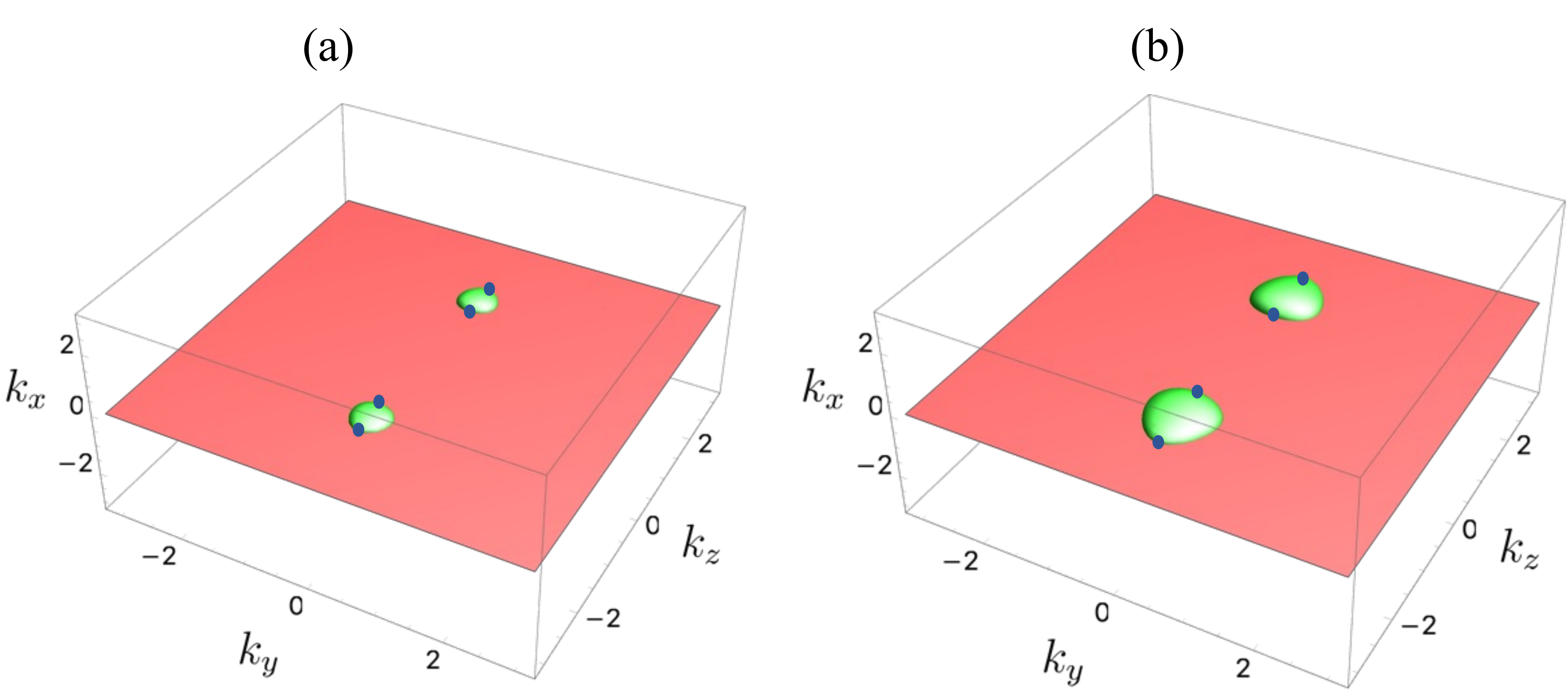}
    \caption{\textbf{Illustration of exceptional contours (ECs) in NH WSM model.} The real (green) and imaginary (red) parts of the exceptional surfaces of the NH WSM are shown for (a) $\gamma=0.3$ and (b) $\gamma=0.5$. The intersection of these two surfaces are the ECs. We note that as we increase the non-Hermitcity strength the size of the ECs increases. Also, four EPs (blue dots) get formed at $k_x=k_y=0$ along $k_z$. The other parameters are chosen to be $m=2$ and $t=r=t_3=1$.}
    \label{fig:BandDia}
\end{figure*}

Recently, NH topological phases have drawn considerable attention of the research community ranging from photonics to condensed matter physics~\cite{rudner2009topological,hu2011absence,esaki2011edge,liang2013topological,malzard2015topologically,san2016majorana,lee2016anomalous,harter2016pt,leykam2017edge,xu2017weyl,feng2017non,el2019dawn,longhi2018parity,ashida2020non,bergholtz2021exceptional,wang2021topological,de2022non,zhang2022review,banerjee2022non}. Dissipation in both classical and quantum mechanical systems is quite common, and this may lead to NH loss and gain. Recent experimental endeavours~\cite{cerjan2019experimental,liu2022experimental,zhao2019non} in controlling dissipation have brought prodigious versatility in the synthesis of NH properties in open classical and quantum systems. In particular, enormous interest has grown towards the topological properties of NH systems, which exhibit unique features absent in their Hermitian counterparts. Strikingly, in NH systems at some particular points of the spectra, the energy eigenvalues and eigenvectors coalesce; in other words, the Hamiltonian describing the system becomes defective. These points are known as EPs~\cite{dembowski2004encircling,heiss2012physics,kato2013perturbation}.

EPs and their intricate structure in different dimensions lead to various kinds of exceptional manifolds, such as lines, rings, surfaces, and complex nexus structures with distinctive electronic excitations and unique Fermi surfaces \cite{xu2017weyl,yoshida2019symmetry,zhang2019experimental,zhou2019exceptional,tang2020exceptional,he2020double,wang2021simulating}. The study of EPs and their underlying spectral topology, as well as their stability in various dimensions in the presence of different kinds of unitary and anti-unitary symmetries, has become an intriguing aspect of non-Hermitian topological phases. Furthermore, higher dimensional exceptional surfaces dubbed as exceptional contours~\cite{cerjan2018effects,yan2021unconventional}, comprised of continuum of EPs, have been recently realized in photonic crystals with the topological charge preserved on the contour~\cite{,cerjan2016exceptional,cerjan2019experimental}. 

Introducing dissipation in a controllable manner in topological phases provides a plethora of brand-new properties, which include EPs with unique spectral degeneracies, skin effects~\cite{yao2018edge,kunst2018biorthogonal,longhi2019probing,li2020critical,kawabata2020higher,yokomizo2021scaling,zhang2022review}, NH topological systems with exotic bulk Fermi arcs~\cite{kozii2017non,zhou2018observation} and Fermi surface topology~\cite{chowdhury2022exceptional}. Novel topological properties of various NH systems in the presence of driving have also been studied~\cite{zhou2019non,zhou2019dynamical,banerjee2020controlling,pan2020non,zhou2021dual,chowdhury2021light,zhou2022driving} in the past few years. The advances along the experimental front have also been remarkable, leading to new applications ranging from topological lasers~\cite{feng2014single,hodaei2015parity} and topo-electrical circuits~\cite{schindler2011experimental,stegmaier2021topological,xiao2019enhanced} to NH transport in driven systems~\cite{zhao2019non}.

In order to establish a bridge between the NH topological systems and their transport properties, a few efforts have been made in the recent years~\cite{philip2018loss,chen2018hall,groenendijk2021universal,wang2022hall,tzortzakakis2021transport,wu2022non,ganguly2022transport}. One such transport property is the Hall conductance. In Hermitian two-dimensional Chern insulators, the Hall conductance is quantized and is given by $\sigma=(e^{2}/h)C$, which is the celebrated Thouless-Kohmoto-Nightingale-den Nijs (TKNN) formula~\cite{thouless1982quantized} for a two-dimensional system. Here $C$ is the Chern number, the topological invariant of the system. However, in the presence of the non-Hermiticity the usual nature of the Hall conductance of the two-dimensional system starts to deviate from the quantized value~\cite{philip2018loss,chen2018hall,groenendijk2021universal,wang2022hall}. However, the nature of the Hall conductance of a three-dimensional system in the presence of non-Hermiticity has not been thoroughly investigated so far. 
 
Our goal in the present manuscript is twofold. We first examine the nature of the non-quantized Hall conductance of an NH WSM with a variation of the non-Hermiticity parameter. The deviation of the Hall conductance from the quantized nature follows a particular pattern, i.e., it remains constant at small momenta, then exhibits a shoulder between a pair of EPs, and eventually becomes zero outside the EPs. Our second goal is to analyze the open boundary condition (OBC) spectra and to determine the topological invariant for this model. As is evident from the literature, the OBC and usual Bloch spectra show differences in the NH cases due to the skin effect~\cite{lee2016anomalous,alvarez2018non,yao2018edge}, where a large number of states get localized at the edge under OBC. This localization of states leads to the violation of the bulk boundary correspondence (BBC)~\cite{yang2022non,lee2016anomalous}. Furthermore, it is important to note that the NH Bloch bands in our model continue to be gap-less, making it ill-defined to determine the usual topological invariant, the Chern number, using the standard Bloch theory prescription. One may alternatively use the non-Bloch theory to compute the Chern numbers by making use of the complex momenta. Thus we provide a complete prescription of framing the problem of breaking of BBC in our system and how to redefine the topological invariant for our system. 

The rest of the paper is organized as follows: In Section~\ref{section:NH_WSM}, we start with the Hamiltonian of the stacked NH Chern insulator and discuss the complex eigenspectra. Next, we discuss the Hall conductance for our model and the deviation of the Hall conductance from the usual quantized pattern is analyzed. In Section~\ref{section:OBC}, the spectra for OBC are analyzed and it is found that the phase transition points are different for the OBC and the Bloch theory. Subsequently, the non-Bloch theory is invoked to show that the momentum phase transition values for OBC and non-Bloch theory match exactly in the GBZ. Finally, we conclude in Section~\ref{section:conclusion}.

\section{Non-Hermitian Weyl semimetal}
\label{section:NH_WSM}

We consider a stack of 2D layers of Chern insulators forming a 3D WSM~\cite{shapourian2016phase}. The Hamiltonian of our system describing the two-band model is

\begin{equation}
    \begin{aligned}
    H(\textbf{k})  & =(t\sin{k_x}+i\gamma)\sigma_x+t\sin{k_y}\sigma_{y}\\ 
    &\qquad+(m-r\cos{k_x}-r\cos{k_y}-t_3\cos{k_z})\sigma_z\label{eqn:1},
    \end{aligned}
\end{equation}

where $t$ and $r$ are the inter-cell hopping amplitudes. Here $m$ is the onsite energy, and $t_3$ is the inter-layer hopping amplitude. We introduce the intra-cell hopping amplitude $i\gamma$ in each 2D layer, which results in non-Hermiticity and thus the Hamiltonian describes an NH WSM.

\begin{figure}
    \centering
    \includegraphics[width=0.98\linewidth]{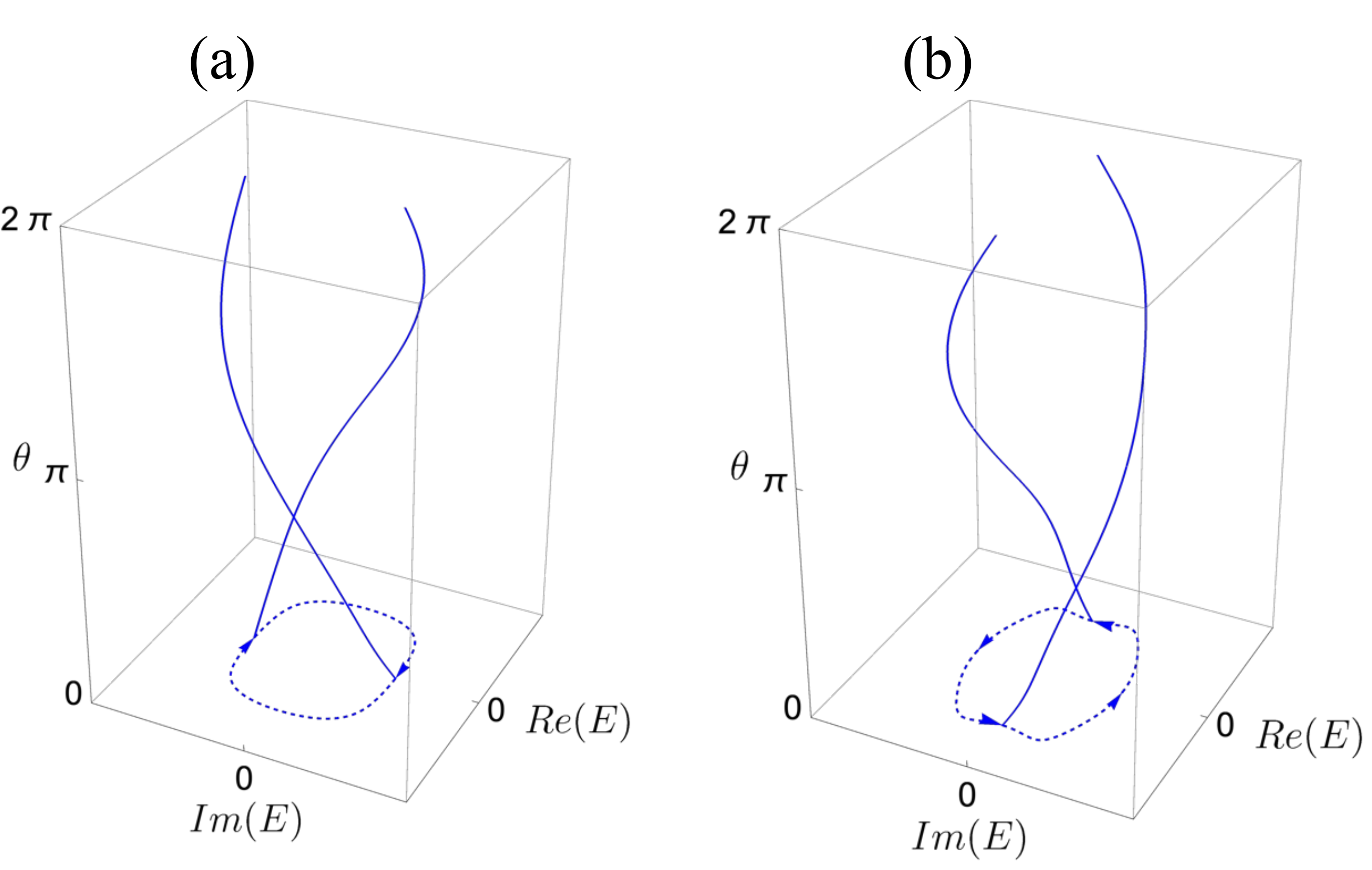}
    \caption{\textbf{Vorticity around EPs.} Vorticity of a pair of EPs when the loop $\Gamma$ encloses them. The plots show the swapping of two energy bands by parameterizing the loop $\Gamma$ using $\theta\in[0,\pi]$. The dashed curves are the projection of the trajectory of the energy bands in the complex plane. The two bands encircle EPs with opposing vorticities as they wrap around one another in opposite (clockwise and anticlockwise) directions in the two plots, (a) and (b). In (a) the contour encloses the EP located at $k_z=k_0^{i}$, and in (b) it encloses the one located at $k_z=k_0^{o}$. The parameters are chosen to be $m=2$, $t=r=t_3=1$, and $\gamma=0.5$.}
    \label{fig:vorticity}
\end{figure}

In the absence of the imaginary term $i\gamma$, the Hermitian Hamiltonian supports a pair of Weyl points when the parameters satisfy the condition $|m-2r|<t_{3}$. The coordinates of the Weyl points are given by $k_x=0,k_y=0,k_z=\pm k_0$ with $k_0=\arccos{(\frac{m-2r}{t_{3}})}$. When $k_z$ lies between $\pm k_0$, it supports a topologically non-trivial phase where the Chern number is 1. However, for $k_z$ outside this range, a topologically trivial phase is obtained where Chern number becomes zero. In most of our analysis, we choose $m=2$ and $t=r=t_3=1$, so that $k_0=\frac{\pi}{2}$.

For our NH system, the energy eigenvalues are given by,

\begin{equation}
\begin{aligned}
\epsilon_{\pm}=\pm[(t\sin{k_x}+i\gamma)^2+(t\sin{k_y})^2+\\
    \quad(m-r\cos{k_x}-r\cos{k_y}-t_3\cos{k_z})^2]^\frac{1}{2}.
\end{aligned}
\end{equation}

A pair of EPs appear corresponding to each Weyl point along the $k_x=k_y=0$ line at $|k_z|=k_0^{i,o}=\arccos{(\frac{m-2r\pm\gamma}{t_{3}})}$. Here $k_0^{i}$ and $k_0^{o}$ are the coordinates of inner and outer EPs, respectively. Thus two pairs of EPs appear when $k_z\in (-\pi,\pi)$. For $k_y\neq 0$, ECs appear in the $k_x=0$ plane. We choose the WSM phase by fixing the parameters and as a result in our NH system, it is possible to adjust the position of the EPs along the $k_z$ direction as well as the sizes of the ECs by tuning the strength of the non-Hermiticity parameter $\gamma$. This has been illustrated and shown in Fig.~\ref{fig:BandDia}. The sizes of the ECs formed in the NH system increases with increasing strength of the non-Hermiticity parameter. Four EPs (blue dots) are formed along $k_x = k_y = 0$ line.  We characterize these EPs topologically by the vorticity associated with them. The vorticity, $v_{mn}$, is defined for a pair of bands as~\cite{shen2018topological}

\begin{equation}
    v_{mn}(\Gamma)=-\frac{1}{2\pi}\oint_{\Gamma}\nabla_{\textbf{k}}~\text{arg}[\epsilon_{m}(\textbf{k})-\epsilon_{n}(\textbf{k})].d\textbf{k},\label{eqn:03}
\end{equation}

where $v_{mn}(\Gamma)$ is the vorticity associated with the bands $m$ and $n$. Here $\Gamma$ represents a closed loop in momentum space. Since eigenvalues for an NH Hamiltonian are in general complex, they can always be written as $\epsilon(\textbf{k})=|\epsilon(\textbf{k})|e^{i\theta_{L}}$, where $\theta_{L}=\arctan{\Big[\frac{\mathrm{Im}(\epsilon)}{\mathrm{Re}(\epsilon)}\Big]}$. Whenever this loop encircles an EP, the vorticity is found to be non-zero. Due to the square root singularity present in the energy eigenvalues, in the complex plane energy bands get exchanged at an EP, and it takes two loops to come back to the initial state. This singularity causes the vorticity to be a half integer. In Fig.~\ref{fig:vorticity}, the vorticity associated with a pair of EPs (for other pair of EPs same follows) located at $k_x=k_y=0$, $k_z=k_z^{i,o}$ in our system is presented. As demonstrated in Fig.~\ref{fig:vorticity}(a) and Fig.~\ref{fig:vorticity}(b), the two bands wind around each other in clockwise and anticlockwise directions, resulting in the vortices of the EPs (at $k_x=k_y=0$, $k_z=k_z^{i,o}$) to be $1/2$ and $-1/2$, respectively. 

Apart from the vorticity, another important topological invariant for NH systems is the winding number. We map out the complete phase diagram enabling NH topological phase transitions by evaluating the winding number. We compute the winding number exploiting the chiral symmetry of the system along $k_y=0$ direction, which has the final form as follows ~\cite{yin2018geometrical,wang2019non} (see Appendix A for a detailed analysis)

\begin{equation}
    \begin{split}
        w & = 
    \begin{cases}
    1, & |k_z|<\sqrt{2(1-\frac{m-2r+\gamma}{t_3})}, \\
    \frac{1}{2}, & \sqrt{2(1-\frac{m-2r+\gamma}{t_3})}<|k_z|<\sqrt{2(1-\frac{m-2r-\gamma}{t_3})}, \\
    0, &|k_z|>\sqrt{2(1-\frac{m-2r-\gamma}{t_3})}.
    \end{cases}
    \end{split}
    \label{winding}
\end{equation}

Such a fractional winding number is a characteristic feature of the NH systems. It is to be noted here that when the closed contour we consider in the calculation of winding number contains two EPs of same winding direction the resulting winding number is $\pm 1.$ However, if the contour encloses a single EP the winding number becomes $\pm 1/2$. Else, the value of the winding number becomes trivial, i.e., 0. In NH systems, it is important to analyze the effect of these EPs on the transport properties of the system. In the next section, we present such an analysis of the Hall conductance for our system.

\section{Hall conductance}

\begin{figure*}
    \centering
    \includegraphics[width=0.8\linewidth]{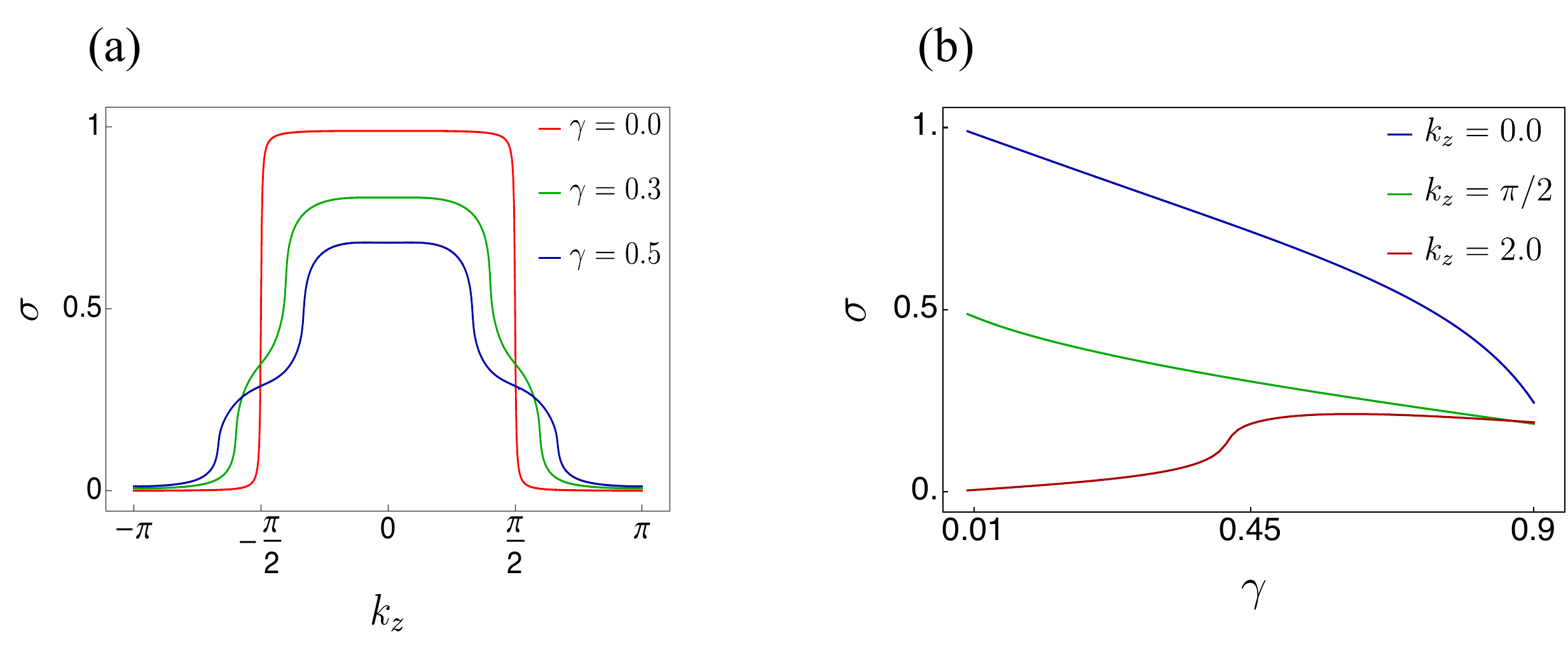}
    \caption{\textbf{Hall Conductance of an NH WSM.} (a) The Hall conductance as a function of $k_z$ given by Eq.~\eqref{eqn:010}. The conductance for three values of $\gamma$ is plotted in units of $\frac{e^2}{h}$. The red line corresponds to the Hermitian case and it is fully quantized. The green and blue curves correspond to NH cases for different strengths of non-Hermiticity, i.e., for $\gamma=0.3$ and $\gamma=0.5$, respectively. They considerably deviate from quantization. In this plot, we identify three different regions between the inner EPs, where the conductance value decreases from 1 but stays constant; between a pair of EPs, where the Hall conductance displays a shoulder shape; and outside the EPs, where it decreases to zero. Here, the nature of the Hall conductance is a direct consequence of the EPs. In plot (b) the Hall conductance is shown as a function of $\gamma$ for three different values of $k_z$ lying in the three different regions identified in (a). The blue curve ($k_z=0$) and the green curve ($k_z=\pi/2$) correspond to the regions where $-k_0^{i}<k_z<k_0^{i}$ and $k_0^{i}<k_z<k_0^{o}$, respectively. In these cases, $\sigma$ decreases with non-Hermiticity strength $\gamma$. The brown curve corresponds to $k_z=2.0$, which lies outside the outer EP for few values of $\gamma$ and for other values of $\gamma$ lies in between the pair of EP. Consequently $\sigma$ remains near zero initially for small values of $\gamma$ and then makes a transition near $\gamma\approx0.43$ to become finite. The parameters are chosen to be $m=2$ and $t=r=t_3=1$. For values of $k_z$ where $\sigma$ was finite in the Hermitian system, $\sigma$ decreases monotonically with increasing non-Hermiticity strength $\gamma$ in the NH system. For other values of $k_z$, initially $\sigma$ remains zero, then it may or may not increase to a finite value depending on the value of $\gamma$.}
    \label{fig:Hallconductance}
\end{figure*}

To study the Hall conductance of our 3D system, let us first illustrate the Hall conductance of a two-band system which is described by a generic two-band Hamiltonian of the form

\begin{equation}
    H(\textbf{k})=d^{0}_{\textbf{k}}\times I+\textbf{d}_{\textbf{k}}\cdot\boldsymbol{\sigma},\label{eqn:05}
\end{equation}

where $d_{j,\textbf{k}}=\text{d}^{R}_{j,\textbf{k}}+i\text{d}^{I}_{j,\textbf{k}}$ ($j=x,y,z$) are in general complex for NH systems. Using linear response theory as presented in Refs.~\onlinecite{chen2018hall,hirsbrunner2019topology,he2020floquet}, the Hall conductance in the $x-y$ plane is written as

\begin{equation}
    \sigma_{xy}=\lim_{\omega\rightarrow 0}\frac{i}{\omega+i0^{+}}[K_{xy}(\omega)-K_{xy}(0)].\label{eqn:06}
\end{equation}

Here $K_{xy}$ is the current-current correlation function and is given by

\begin{equation}
    K_{xy}(\omega)=\sum_{\textbf{k}}\int\frac{d\epsilon d\epsilon'}{\pi^{2}}\frac{n_{F}(\epsilon')-n_{F}(\epsilon)}{\epsilon'-\epsilon+\omega+i0^{+}}\text{Tr}(\hat{J_{x}}A(\epsilon)\hat{J_{y}}A(\epsilon')),\label{eqn:07}
\end{equation}

with $n_{F}(\epsilon)$ as the Fermi distribution function at temperature $T$ and 

\begin{align}
\hat{J}_{x,y}=\frac{\partial {\rm Re}H(\textbf{k})}{\partial k_{x,y}},
\end{align}

denotes the current operator. Here $ A(\epsilon)= {\rm Im Tr}\left[\hat{G}^{R}\right]$ is the spectral function and $\hat{G}^{R}$ is the retarded Green's function defined as

\begin{equation}
\hat{G}^{R}=\sum_{\alpha=\pm,\textbf{k}}\frac{\ket{\phi_{\alpha,\textbf{k}}^{R}}\bra{\phi_{\alpha,\textbf{k}}^{L}}}{\epsilon-\epsilon^{\alpha}_\textbf{k}}.\label{eqn:9}
\end{equation}

Here, $\alpha=\pm$ denotes the two bands of the Hamiltonian. Further, $\ket{\phi_{\alpha,\textbf{k}}^{R(L)}}$ is the right (left) eigenvector of the Hamiltonian. The Hall conductance for such a generic two-band system at zero temperature is found to be~\cite{chen2018hall}

\begin{equation}
    \sigma_{xy}=\sum_{\textbf{k}}\frac{\Omega_{xy}(\textbf{k})+\Omega^{*}_{xy}(\textbf{k})}{2}\times\nu(\textbf{k}),\label{eqn:010}
\end{equation}

where

\begin{equation}
\Omega_{xy}(\textbf{k})=\textbf{d}_{\textbf{k}}.(\partial_{k_{x}}\textbf{d}_{\textbf{k}}\times\partial_{k_{y}}\textbf{d}_{\textbf{k}})/\epsilon(\textbf{k})^{3},
\end{equation}

and

\begin{equation}
\nu(\textbf{k})=2\tan^{-1}(\text{Re}~\epsilon(\textbf{k})/{\rm Im}~\epsilon(\textbf{k}))/\pi.
\end{equation}

Here, $\Omega_{xy}(\textbf{k})$ is the Berry curvature and the term $\nu(\textbf{k})$ captures the effect of non-Hermiticity on the Hall conductance. For our model, we have
$\textbf{d}_{\rm k}=(t\sin{k_x}+i\gamma,t \sin{k_y},(m-r\cos{k_x}-r\cos{k_y}-t_3\cos{k_z}))$. 

\begin{figure}
    \centering
    \includegraphics[width=0.8\linewidth]{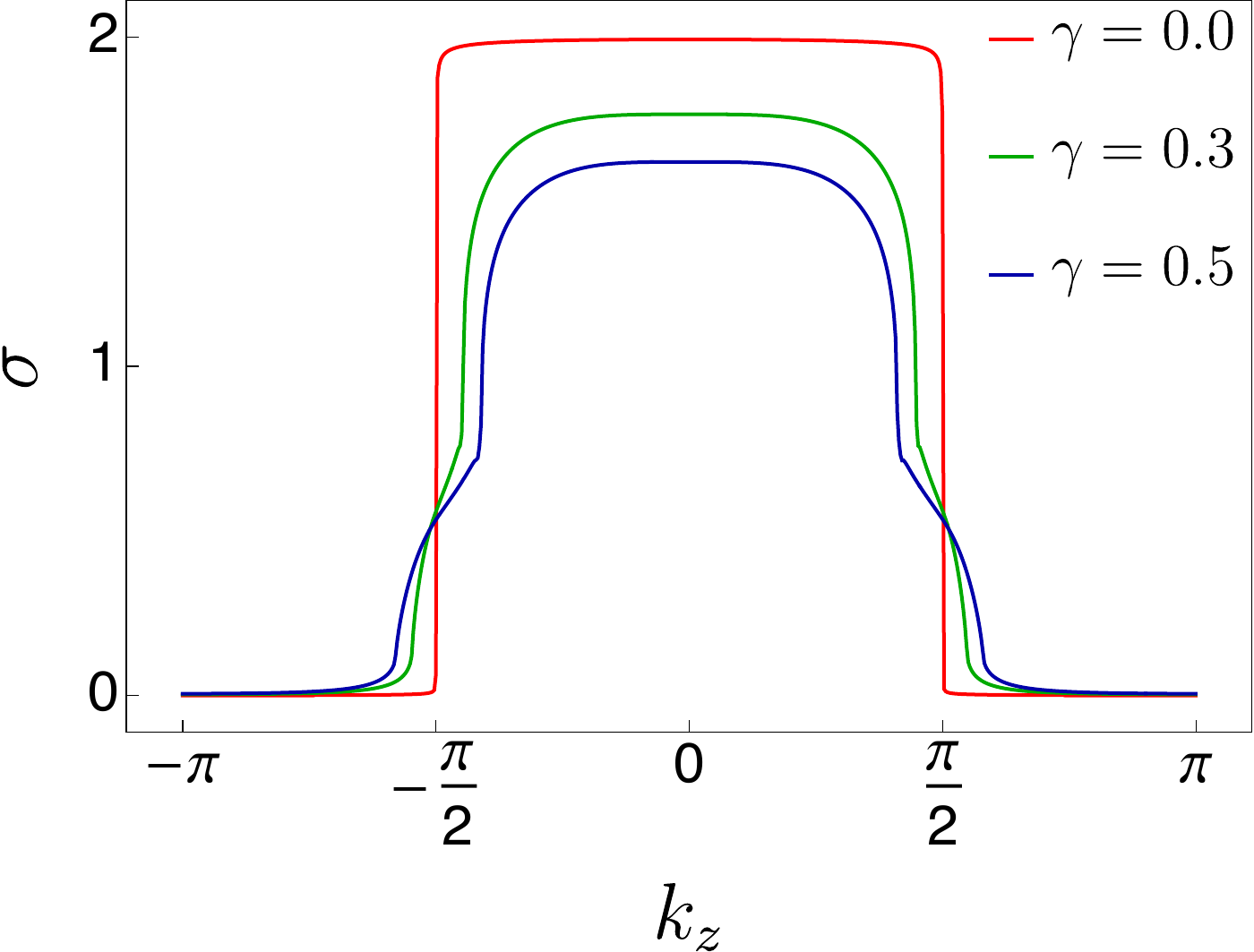}
    \caption{\textbf{Hall conductance of an NH double WSM.} The Hall conductance (in units of $\frac{e^2}{h}$) of the non-Hermitian Stack of Chern insulator with $C=2$ is presented as a function of $k_z$ for different values of $\gamma$. The red solid curve represents the Hermitian case when $\gamma=0$. The green and blue solid curves represent non-Hermitian cases where the $\gamma$ takes values $0.3$ and $0.5$, respectively. The other parameters are chosen to be $m=2$ and $t=r=t_3=1$. The Hall conductance for the NH double WSM varies with momenta $k_z$ in a similar manner to the NH WSM -- it remains nearly constant between two inner EPs, exhibits a shoulder in between a pair of EPs and then falls to zero. The Hall conductance starts deviating from the maximum value $\sigma=2$, indicative of an NH double WSM arising from the Chern insulator $(C=2)$.}
    \label{fig:Hall_C2}
\end{figure}

\begin{figure}
\begin{subfigure}
  \centering
  \includegraphics[width=.8\linewidth]{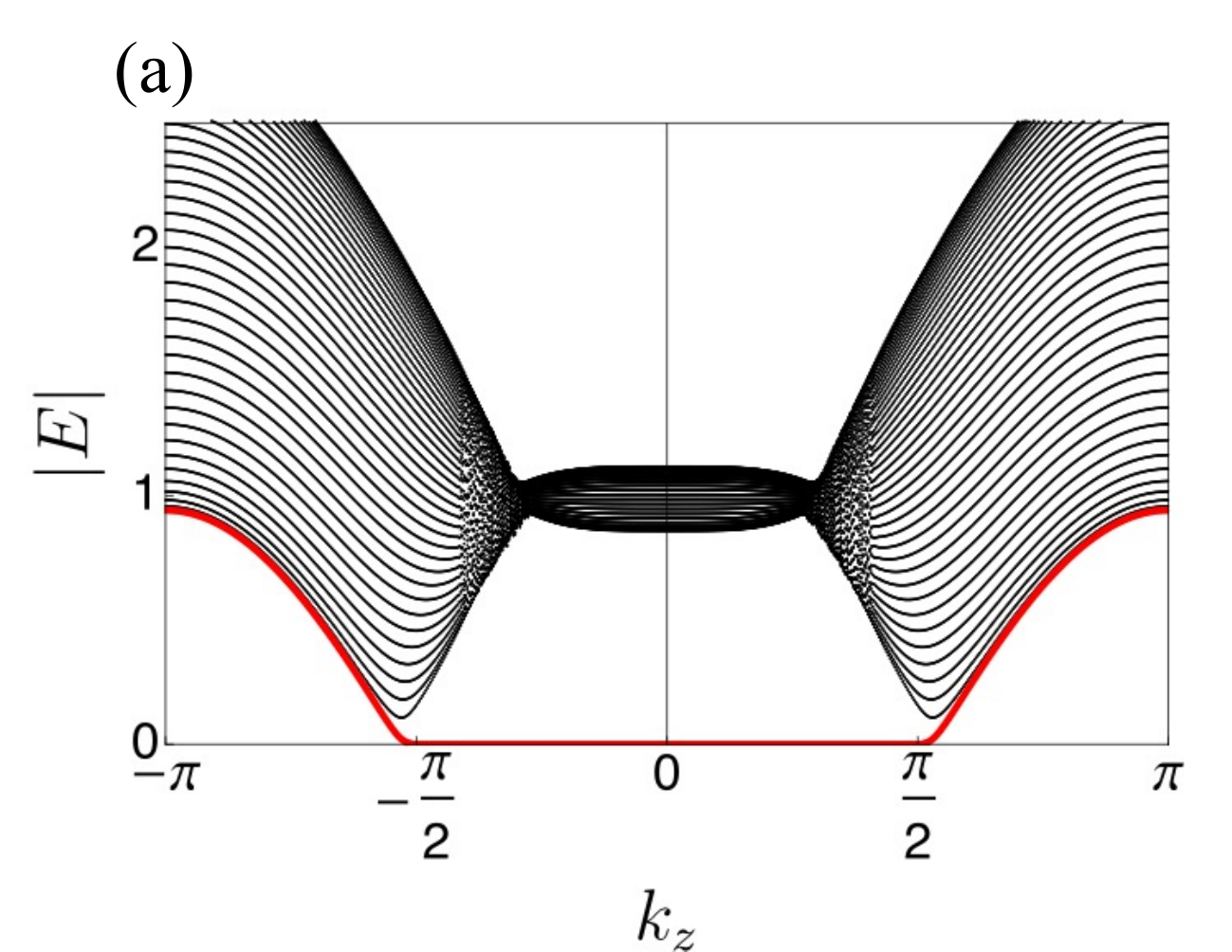}  
  
\end{subfigure}

\begin{subfigure}
  \centering
  \includegraphics[width=.8\linewidth]{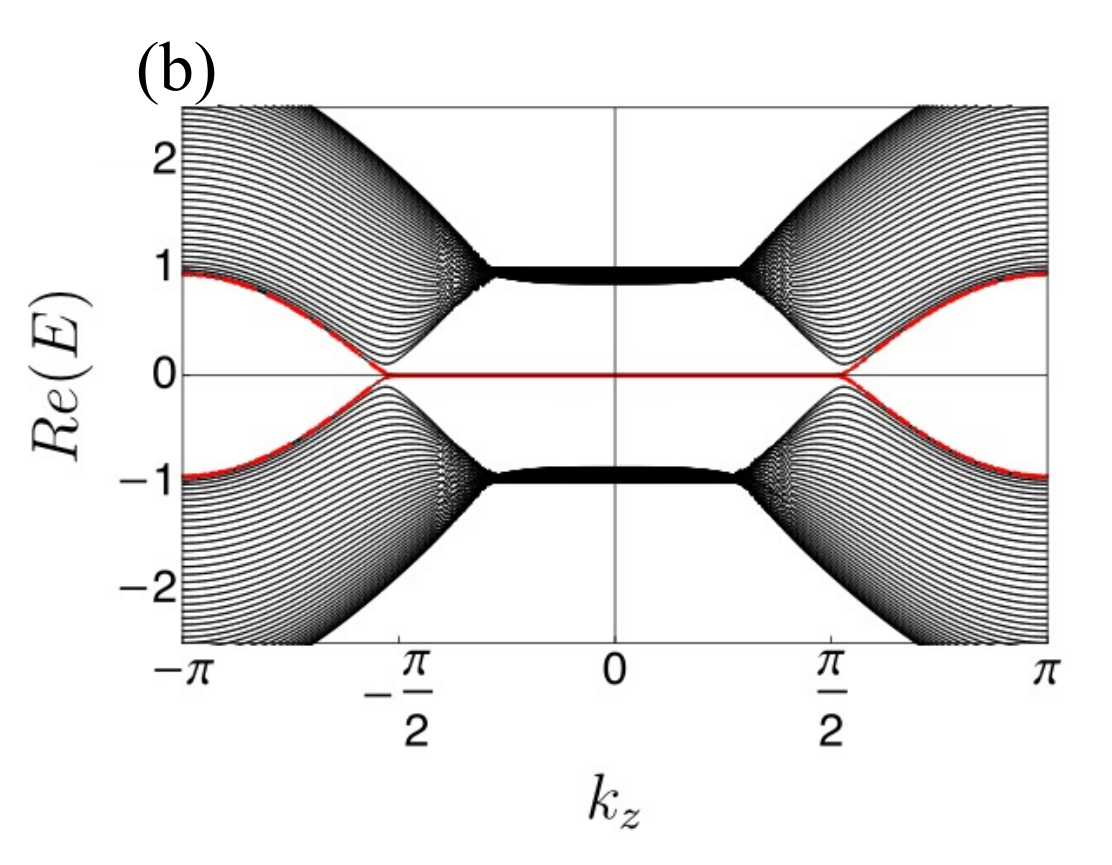}  
\end{subfigure}

\begin{subfigure}
  \centering
  \includegraphics[width=.84\linewidth]{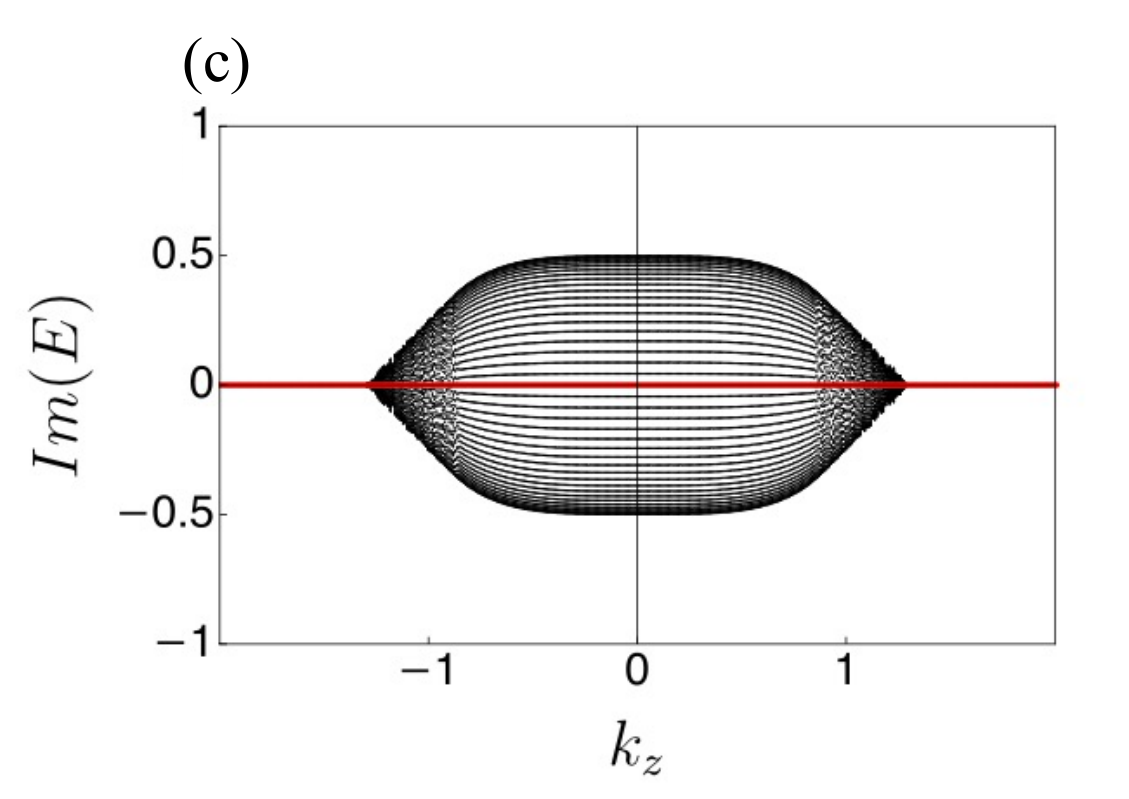}
  
\end{subfigure}
\caption{\textbf{Open boundary spectra for WSM model}. Numerically obtained spectra of the system as a function of $k_z$ for $N=42$ unit cells, when the system has OBCs along the $x$-direction. (a)-(c) are the absolute, real and imaginary parts of the energy spectra. The parameters are chosen to be $m=2$, $t=r=t_3=1$, and $\gamma=0.5$. Here, the zero modes are shown in red. Using OBC we find that the true topological transition point is at $|k_z|=\arccos{(-\frac{\gamma^2}{2})}$.} 
\label{fig:Edgemodes}
\end{figure}

In Fig.~\ref{fig:Hallconductance}(a), the Hall conductance for various strengths of the NH parameter $\gamma$ is presented. For $\gamma=0$, i.e., in the Hermitian case, the Hall conductance remains quantized between the two Weyl nodes, which is represented by the solid red curve in Fig.~\ref{fig:Hallconductance}(a). As we allow $\gamma$ to be non-zero, $\sigma_{xy}$ starts deviating from the quantized value and causes a drop in the maximum value of the Hall conductance. The value of the conductance for a particular $\gamma$ remains constant between two inner EPs. It exhibits a shoulder between a pair of EPs before finally diminishing to zero outside the EPs. The presence of the term $\nu(\textbf{k})$ in the Hall conductance expression (see Eq.~\ref{eqn:010}) causes the deviation from quantization. Physically, the finite imaginary part of the spectra introduces a finite lifetime corresponding to each carrier. The carriers having momenta in $z$-direction in the range $-k_{0}^{i}<k_{z}<k_{0}^{i}$ contribute most in the Hall conductance. In this region, the Hall conductance gets suppressed with increasing $\gamma$ value due to the decreasing lifetime of the carriers. Furthermore, the carriers having only momenta $k_z$ in the region between $(|k_{0}^{i}|,|k_{0}^{o}|)$ contribute in the Hall conductance. Interestingly, the nature of the Hall conductance in this region owes to the fact that a Weyl point in the Hermitian system gives rise to a pair of EPs in the NH system. The presence of EPs leads to short-lived low-lying excitations in this region, resulting in shoulder-like behaviour in the Hall conductance. The carriers that have momenta $|k_z|>k_0^{o}$ do not contribute to the Hall conductance of the system. In Fig.~\ref{fig:Hallconductance}(b), the Hall conductance is presented as a function of $\gamma$ for different values of $k_z$. We choose $\gamma$ in a way that $k_z=0$ always lies in the region where $-k_0^{i}<k_z<k_0^{i}$; thus for $k_z=0$ the Hall conductance $\sigma$ decreases with $\gamma$ from the value of one. Next, we choose $k_z=\pi/2$ -- it is the position of the Weyl point in the parent Hermitian system -- which always lies between a pair of EPs in the NH system. We find that $\sigma$ again decreases with $\gamma$. Finally, when we set $k_z=2.0$, the strength of $\gamma$ determines whether $k_0^{o}<k_z(=2)$ or $k_0^{i}<k_z(=2)<k_0^{o}$. Here $\sigma$ remains zero in the first case and becomes finite in the second case. 

Apart from the stacked $C=1$ Chern insulator case, resulting in a WSM with topological charge 1, we have also analyzed the Hall conductance of a stacked $C=2$ Chern insulator. The Hamiltonian of a WSM formed by stacking 2D Chern insulators of Chern number $C=2$ with intra-cell hopping amplitude $i\gamma$ is written as

\begin{equation}
    \begin{aligned}
    H(\textbf{k})  & =(t\cos{k_x}-t\cos{k_{y}}+i\gamma)\sigma_x+2t\sin{k_x}\sin{k_y}\sigma_{y}\\ 
    &\qquad+(m-r\cos{k_x}-r\cos{k_y}-t_3\cos{k_z})\sigma_z.\label{eqn:dWSM2}
    \end{aligned}
\end{equation}

As long as we are in the parameter regime where $|m-2r|<t_3$, the parent Hermitian double WSM hosts two Weyl points at $(k_x=k_y=0$ and $k_z=\pm\cos{(\frac{m-2r}{t_{3}})}$. The Berry charge for these Weyl points is $\pm 2$. Each Weyl point splits into a pair of EPs when $\gamma \neq 0$. Calculating in the same manner as described above, we find that the Hall conductance for this model behaves similar to NH WSM with topological charge $\pm 1$, as shown in Fig.~\ref{fig:Hall_C2}. The value of the Hall conductance decreases with increasing strength of non-Hermiticity. The similarity between the nature of the Hall conductance of an NH double WSM ($C=2$) and an NH WSM ($C=1$) indicates the generality of our findings. 

We note that the term $\nu(\mathbf{k})$ in Eq.~\ref{eqn:010}, manifesting in the non-Hermitian generalization of the TKNN formula, leads to the lifting of quantization of the Hall conductance even in the presence of a fractional topological index signalling the non-Hermitian phase transitions (see Appendix A for a discussion). Nonetheless, the Bloch theory dramatically fails to match the open boundary edge modes information resulting in broken bulk-boundary correspondence (BBC) ~\cite{yao2018edge,yang2020non,helbig2020generalized,xiao2020non2}. Consequently, the direct mapping between Hall conductance and the existing edge mode transport still needs to be explored. This requires us to restructure the framework, which quantifies the behaviour of edge transport over the GBZ and restores the BBC.

\section{Open Boundary Condition and Non-Bloch theory}
\label{section:OBC}

Having discussed the Bloch band properties of the model using its complex eigenvalue spectrum and its transport properties, we next discuss the OBC and the non-Bloch theory. We find that, strikingly, our system encounters the NH skin effect (NHSE). This dictates that a macroscopically large number of states are exponentially localized at the edge under OBC, leading to the violation of the celebrated BBC~\cite{yang2022non,lee2016anomalous}. The condition for obtaining zero modes under OBC in such systems was also discovered to be different from the periodic boundary conditions in the Bloch band theory framework. These consequences suggest a re-examination of the topological invariants in the GBZ to characterize their topology in terms of open boundary modes. Next, our analysis goes as follows. First, we numerically investigate the spectral topology under OBC. Then we employ the non-Bloch theory to find the topological zero modes enabling topological phase transitions. Finally, we characterize the edge transport and analyze the Hall conductance defined over GBZ in terms of the exceptional Weyl points and boundary states.
 
Since in our system, the degeneracies in the band diagrams are found to be in the $k_x=0$ plane, therefore, we consider the system to be open along the $x$-direction and treat $k_y$, $k_z$ as parameters. We then investigate the OBC spectra for $N=42$ unit cells with the parameters fixed at $m=2$, $r=t=t_3=1$, and $\gamma=0.5$ in Fig.~\ref{fig:Edgemodes}, where the absolute, real and imaginary parts of energy are plotted as a function of $k_z$. The zero energy (both real and imaginary) edge modes are shown in red. In our system, the NHSE gives rise to hybridization between the edge mode and bulk skin modes that possess finite imaginary energy components, leading to their participation in transport. As a result, the Hall conductance deviates from quantization. Although, purely real zero-energy edge modes (shown in red) retain a non-decaying current enabling edge transport.

\begin{figure}
    \centering
    \includegraphics[width=0.9\linewidth]{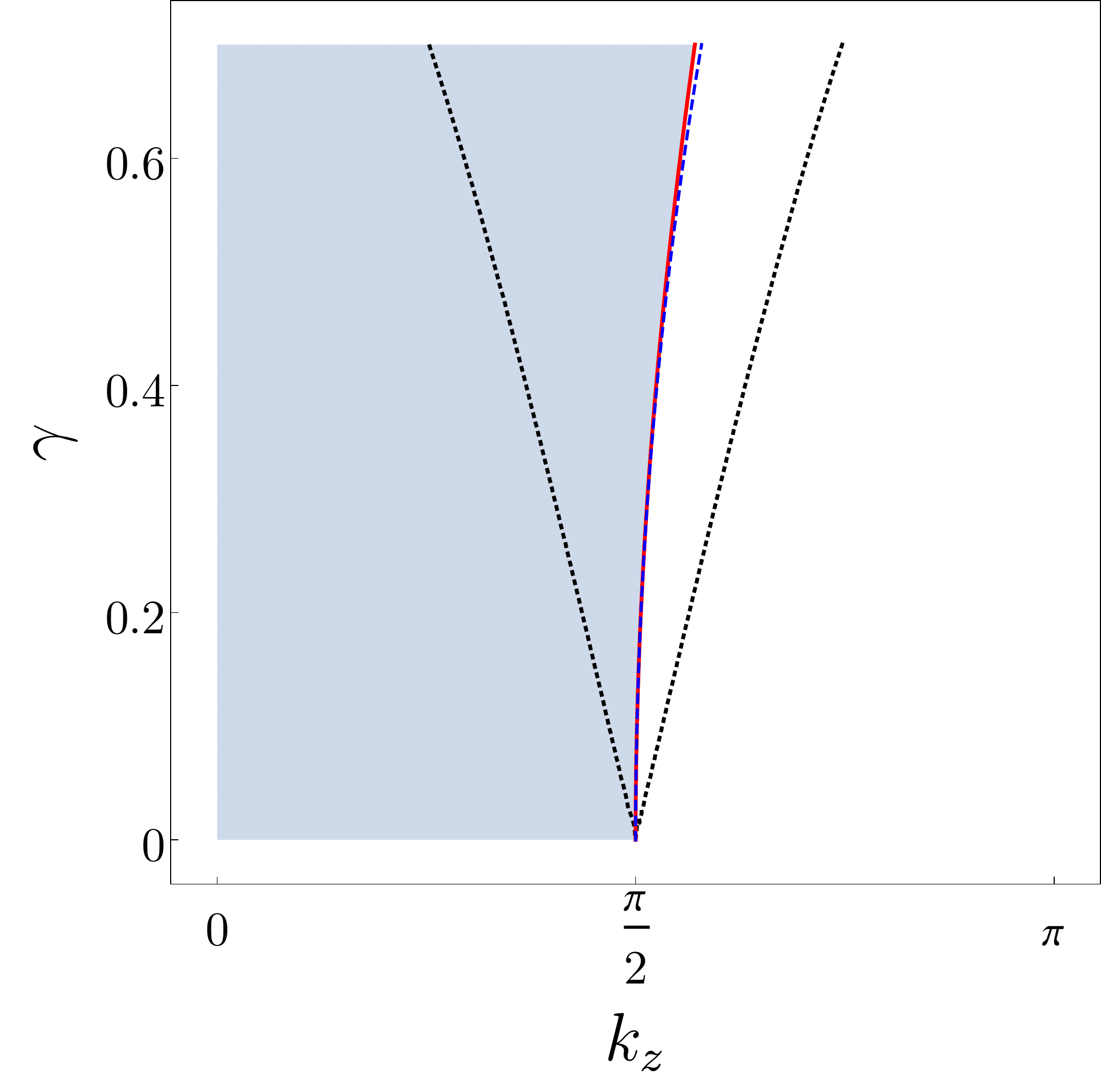}
    \caption{\textbf{Topological phase diagram.} The black dotted lines are the phase boundaries ($k_z=k_z^{i}$ and $k_z=k_z^{o}$) derived using the Bloch model, where for $k_z^{i}<k_z<k_z^{o}$ the system remains gapless. The red line shows the phase boundary obtained using OBC and the blue dashed line corresponds to the non-Bloch continuum model. We note that the Bloch phase boundary significantly differs from the OBC spectra, which invalidates the conventional BBC. This mismatch suggests a non-Bloch framework to characterize their topological-trivial phase transitions. As we see in the plot, the blue dashed line obtained from the non-Bloch theory nicely captures the topological phase transitions. The edge modes will be found in the shaded region. Therefore, this region is the topologically non-trivial phase. The other parameters are chosen to be $m=2$ and $t=r=t_3=1$.}
    \label{fig:Phasediagram}
\end{figure}

Next, we study the non-Bloch band theory which restores the BBC, following the formalism derived in Ref.~\onlinecite{yao2018non}. First, we consider the low energy continuum model derived from our parent Hamiltonian in Eq.~\eqref{eqn:1} by setting $\sin{k_{x,y}}\approx k_{x,y}$, $\cos{k_{x,y}}\approx 1-\frac{k_{x,y}^2}{2}$ and treating $k_z$ as parameter

\begin{equation}
    \begin{aligned}
    H(\textbf{k})& = (tk_x+i\gamma)\sigma_{x}+tk_y\sigma_{y}\\
    &\qquad+(m-r(1-\frac{k_x^{2}}{2})-r(1-\frac{k_y^{2}}{2})-t_3\cos{ k_z})\sigma_z.\label{eqn:014}
    \end{aligned}
\end{equation}

We consider the wave vector to be complex-valued as

\begin{equation}
    \textbf{k}\rightarrow\Tilde{\textbf{k}}^{r}+\Tilde{\textbf{k}}^{i},\label{eqn:15}
\end{equation}

where $\Tilde{\textbf{k}}^{r}=(k_x,k_y,k_z)$ and $\Tilde{\textbf{k}}^{i}=(\frac{\gamma}{t},0,0)$. We obtain the non-Bloch Hamiltonian for our model to be

\begin{equation}
   \begin{aligned}
        H(\Tilde{\textbf{k}})&=t\Tilde{k}_{x}\sigma_x+t\Tilde{k}_{y}\sigma_y\\
        &\qquad+\left[\Tilde{m}+r\left(\frac{\Tilde{k}_x^{2}+\Tilde{k}_y^{2}}{2}\right)-i\frac{r\gamma}{t}\Tilde{k}_x\right]\sigma_z.\label{eqn:16}
    \end{aligned}
\end{equation}

Here $\Tilde{m}=m-2r-\frac{r\gamma^2}{2t^2}-t_3\cos{k_z}$. The wave vector $\Tilde{\textbf{k}}$ lies in the GBZ in the complex plane just as $\textbf{k}$ lies in the conventional Brillouin zone in the real plane. The non-Bloch Chern number $\tilde{C}$ can be determined from the Hamiltonian (Eq.~\eqref{eqn:16}) using the usual prescription~\cite{yang2022non,yang2022non}. $\tilde{C}$ is $1$ when $\Tilde{m}<0$ and $0$ when $\Tilde{m}>0$. Thus, using non-Bloch theory we can determine the condition for the our system to undergo a phase transition from topologically non-trivial phase to normal insulator phase. This is given by

\begin{equation}
    k_z=\pm\arccos{\frac{m-2r-\frac{r\gamma^2}{2t^2}}{t_3}}.\label{eqn:17}
\end{equation}

Substituting the values of the parameter that we fixed initially, we finally have $k_z=k_{z}^{c}=\arccos({-\frac{\gamma^2}{2}})$. Notably, this is different from the topological transition point derived using Bloch band theory but exactly matches with the condition obtained using the OBC spectra. This is illustrated in Fig.~\ref{fig:Phasediagram}, where the phase diagram for our model is presented. Since the Hamiltonian is symmetric in $k_z$, we only choose positive values of $k_z$ to plot the phase diagram. In Fig.~\ref{fig:Phasediagram}, the shaded region denotes the topologically non-trivial phase $(\tilde{C}=1)$, whereas the black dotted lines are the phase boundaries found from the Bloch continuum model. The overlapping red solid and blue dashed lines are the phase boundaries derived using the OBC computations and the non-Bloch continuum model, respectively. This results in the restoration of BBC in our NH system. 

The energy bands of the non-Bloch Hamiltonian touch at $\pm k_{z}^{c}$ at discrete points, leading to Weyl EPs under PBC. The GBZ allows us to evaluate the total Hall conductance. We find it as~\cite{shapourian2016phase,burkov2011weyl}

\begin{equation}
    \sigma_{xy}=\frac{e^2}{4\pi h}\int dk_z \tilde{C}=\frac{\text{e}^{2}k_{z}^{c}}{\pi h},\label{eqn:18}
\end{equation}

where $\tilde{C}$ is the non-Hermitian Chern number defined over the GBZ. Therefore, the Hall conductance in WSM is proportional to the distance between the two NH Weyl points. This parallels the Hermitian case, but notably one requires the formulation of the GBZ to obtain this.

\section{Conclusions}
\label{section:conclusion}

In this work, we explored the transport properties of a 3D NH WSM formed via stacking 2D NH Chern insulators. We discover that in such an NH system the Hall conductance deviates from the quantized value and exhibits a shoulder-like character, which we interpret as a consequence of the existence of pairs of EPs. Notably, the finite imaginary part of the energy introduces a finite lifetime of the carriers. The carriers having momenta in $z$-direction in the range $-k_{0}^{i}<k_{z}<k_{0}^{i}$ contribute the most to the Hall conductance, where the Hall conductance varies inversely with $\gamma$ due to the short lifetime of the carriers. However, when the $k_z$ value of the carriers lie between $(|k_{0}^{i}|,|k_{0}^{o}|)$, there is a finite contribution of the carriers to the Hall conductance. The OBC and the usual Bloch theory disagree for such NH systems and we show that the transition between the topologically non-trivial phase to the trivial phase occurs for different $k_z$ values for OBC and Bloch theory. This discrepancy was solved by using the GBZ and complex momentum values, i.e., the non-Bloch theory. We presented the Hall conductance evaluated over the GBZ and connected it to the separation between the Weyl nodes. We hope our findings stimulate further exploration of unusual transport properties of NH systems.

\section{Acknowledgements}
S.D. and A.B. are supported by Prime Minister's Research Fellowship (PMRF). D.C. acknowledges financial support from DST (Project No. SR/WOS-A/PM-52/2019). S.D. thanks A. Ghosh and S. Basu for their help. A. N. acknowledges support from the startup grant at Indian Institute of Science (SG/MHRD-19-0001).

\appendix
\section{Calculation of winding number}

To calculate another topological invariant, namely the winding number, we consider the low energy continuum model of our system in the $k_y=0$ plane. The Hamiltonian has chiral symmetry in this plane, i.e., $\sigma_yH(\textbf{k})\sigma_y=-H(\textbf{k})$. Thus, by treating $k_x$ and $k_z$ as parameters and considering 1D chains along $k_x$ direction, the winding number can be calculated as~\cite{yin2018geometrical}

\begin{equation}
    w=\frac{1}{2\pi}\int_{-\infty}^{\infty}dk_x\partial_{k_x}\phi,\label{eqn:3}
\end{equation}

where $\phi=\arctan{(d_z/d_x)}$, $d_z$, $d_x$ represent the coefficients of $\sigma_z$, $\sigma_x$ terms in $H$, respectively. Now, the low energy continuum model of our system in the $k_y=0$ plane is given by

\begin{equation}
    H(\textbf{k})\approx (tk_x+i\gamma)\sigma_{x}+(m-r(1-\frac{k_x^{2}}{2})-r-t_3(1-\frac{ k_z^{2}}{2}))\sigma_z.\label{eqn:4}
\end{equation}

Therefore, the winding angle becomes
\begin{equation}
    \begin{aligned}
        \phi &=\arctan{(\frac{d_z}{d_x})}\\
        &=\arctan{((m-r(1-\frac{k_x^{2}}{2})-r-t_3(1-\frac{ k_z^{2}}{2}))/tk_x+i\gamma)}.
    \end{aligned}
\end{equation}
The presence of the imaginary term in the Hamiltonian makes the winding angle, $\phi$ complex. Thus, $\phi=\phi_R+i\phi_I$, where $\phi_R$, $\phi_I$ are the real and imaginary parts of the winding angle, respectively. Following the method outlined in Ref.~\onlinecite{yin2018geometrical}, first the values of $\phi$ at the limiting values of $k_x \rightarrow{\pm\infty}$ are evaluated as,

\begin{equation}
    \phi_{k_x\rightarrow{\pm\infty}}=\arctan\left(\frac{d_z}{d_x}\right)_{k_x\rightarrow{\pm\infty}}=\pm\frac{\pi}{2},\label{eqn:5}
\end{equation}

which are purely real. Now using the relation

\begin{equation}
    \text{e}^{2i\phi}=\frac{\cos\phi+i\sin\phi}{\cos\phi-i\sin\phi}=\frac{1+i\tan\phi}{1-i\tan\phi}=\frac{d_x+id_z}{d_x-id_z},\label{eqn:6}
\end{equation}

one can obtain that the $\phi_R$ and $\phi_I$ are related to the phase and amplitude as

\begin{equation}
    \text{e}^{-2\phi_I}=\left|\frac{d_x+id_z}{d_x-id_z}\right|,\label{eqn:7}
\end{equation}

and 

\begin{equation}
    \text{e}^{2i\phi_R}=\frac{d_x+id_z}{d_x-id_z} \Biggm/ \left|\frac{d_x+id_z}{d_x-id_z}\right|.\label{eqn:8}
\end{equation}

\begin{figure}
    \centering
    \includegraphics[width=0.8\linewidth]{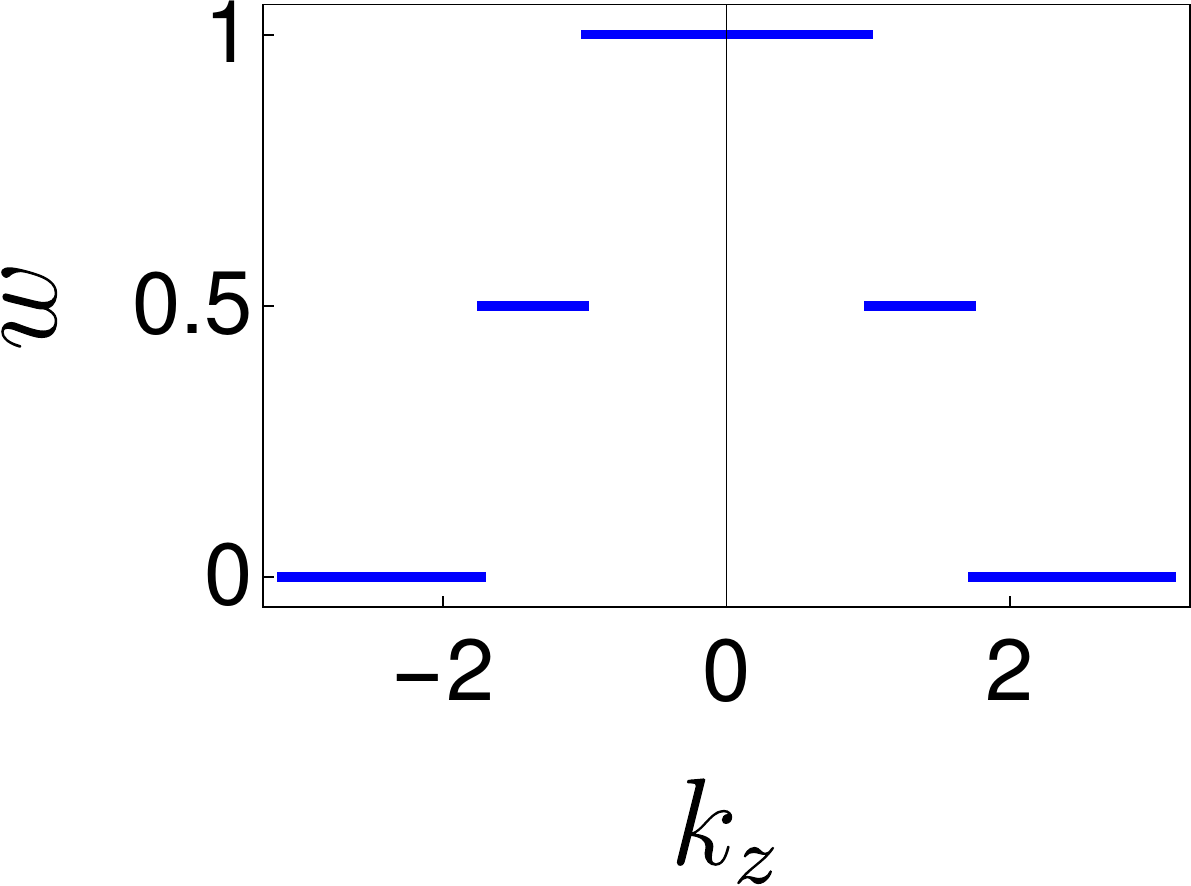}
    \caption{\textbf{Winding number.} A plot of the winding number as a function of $k_z$ calculated considering the low energy model. The parameters are chosen to be $m=2$, $t=r=t_3=1$, and $\gamma=0.5$. There are three distinct topological phases with winding numbers $1,~1/2,~0$ respectively. Since $\partial_{k_{x}}\phi$ (where $\phi$ is the winding angle) is same at the two limits of $k_{x}$, i.e., at $k_{x}=\pm\pi/2$, the integral contour in the calculation of winding number can be constructed as a loop. The fractional value of winding number $1/2$ arises when the loop encloses only one EP. Further $w$ becomes $1$ when two EPs are enclosed and zero when the contour does not enclose any EP.}
    \label{fig:WindingNumber}
\end{figure}

The imaginary part of the winding angle $\phi_I$ is found to be a real continuous function of $k_x$, thus

\begin{equation}
    \int_{-\infty}^{\infty}dk_x\partial_{k_x}\phi_I=\phi_I|_{k_x\rightarrow{\infty}}-\phi_I|_{k_x\rightarrow{-\infty}}=0.\label{eqn:19}
\end{equation}

Further, one has

\begin{equation}
    \tan(2\phi_R)=\frac{\text{Im}(\frac{d_x+id_z}{d_x-id_z})}{\text{Re}(\frac{d_x+id_z}{d_x-id_z})}.\label{eqn:11}
\end{equation}

Also, we have

\begin{equation}
    \tan(2\phi_R)=\tan(\phi_A+\phi_B),\label{eqn:12}
\end{equation}

with the real angles $\phi_A$, $\phi_B$ defined as~\cite{yin2018geometrical,wang2019non}

\begin{equation}
    \begin{aligned}
        \tan\phi_A & =\frac{\text{Re}(d_z)+\text{Im}(d_x)}{\text{Re}(d_z)-\text{Im}(d_x)} \\
        & =\frac{m-2r+r\frac{k_x^{2}}{2}-t_3(1-\frac{k_z^{2}}{2})+\gamma}{tk_x},  \\
        \tan\phi_B & =\frac{\text{Re}(d_z)-\text{Im}(d_x)}{\text{Re}(d_z)+\text{Im}(d_x)} \\
        & =\frac{m-2r+r\frac{k_x^{2}}{2}-t_3(1-\frac{k_z^{2}}{2})-\gamma}{tk_x}.\label{eqn:13}
    \end{aligned}
\end{equation}

This leads to

\begin{equation}
    \phi_R=n\pi+\frac{1}{2}(\phi_A+\phi_B).\label{eqn:14}
\end{equation}

Since $\phi_A$ and $\phi_B$ both have discontinuity at $k_x=0$, one obtains 

\begin{equation}
    \begin{aligned}
    \phi_A(k_x\rightarrow{0^{\pm}})=\pm\frac{\pi}{2}\text{sgn}(m-2r-t_3(1-\frac{k_z^{2}}{2})+\gamma), \\
    \phi_B(k_x\rightarrow{0^{\pm}})=\pm\frac{\pi}{2}\text{sgn}(m-2r-t_3(1-\frac{k_z^{2}}{2})-\gamma).
    \end{aligned}
\end{equation}

Furthermore

\begin{equation}
    \phi_A(k_x\rightarrow{\pm\infty})=\phi_B(k_x\rightarrow{\pm\infty})=\pm\frac{\pi}{2}.
\end{equation}

Writing $(m-2r-t_3(1-\frac{k_z^{2}}{2})+\gamma)=\alpha_1$ and $(m-2r-t_3(1-\frac{k_z^{2}}{2})-\gamma)=\alpha_2$, we get

\begin{equation}
    \begin{split}
        w & =\frac{1}{2\pi}\int_{-\infty}^{\infty}dk_x\partial_{k_x}\phi_R \\
    & = \frac{1}{4\pi}\int_{-\infty}^{\infty}dk_x\partial_{k_x}(\phi_A+\phi_B) \\
    & = \frac{1}{4\pi}((\phi_A|^{\infty}_{0^{+}}+\phi_A|^{0^{-}}_{-\infty})+(\phi_B|^{\infty}_{0^{+}}+\phi_B|^{0^{-}}_{-\infty})) \\
    & = -\frac{1}{2}-\frac{\text{sgn}(\alpha_1)+\text{sgn}(\alpha_2))}{4}.
    \end{split}
\end{equation}

Now substituting the values $\alpha_1$ and $\alpha_2$ and considering different values of $k_z$ the winding number $w$ of the system is as follows

\begin{equation}
    w=\begin{cases}
        1, & |k_z|<\sqrt{2(1-\frac{m-2r+\gamma}{t_3})} \\
    \frac{1}{2}, & \sqrt{2(1-\frac{m-2r+\gamma}{t_3})}<|k_z|<\sqrt{2(1-\frac{m-2r-\gamma}{t_3})} \\
    0, &|k_z|>\sqrt{2(1-\frac{m-2r-\gamma}{t_3})}.
    \end{cases}
\end{equation}

In Fig.~\ref{fig:WindingNumber}, the variation of the winding number, $w$, as a function of $k_z$ is shown. There are three distinct topological phases along the $k_z$ direction corresponding to the three different values of winding number $1$ (enclosing two EPs), $1/2$ (enclosing one EP), and $0$ (not enclosing any EP).

\bibliography{references.bib}

\end{document}